\documentclass[12pt,preprint]{aastex}

\shorttitle{Nuclear vs. Circumnuclear Activity}
\shortauthors{D\'{\i}az, Ag\"uero \& Dottori}

\begin{document}

\title{NUCLEAR vs. CIRCUMNUCLEAR ACTIVITY}

\author{Rub\'en J. D\'{\i}az\altaffilmark{1}, Mar\'{\i}a P. Ag\"uero\altaffilmark{1,2} \& Horacio A. Dottori\altaffilmark{3}}

\altaffiltext{1}{Observatorio Astron\'omico de C\'ordoba, UNC,
Laprida 854, 5000 C\'ordoba, Argentina} \altaffiltext{2}{SeCyT,
Universidad Nacional de C\'ordoba, Argentina}
\altaffiltext{3}{Instituto de F\'{\i}sica -- Universidade Federal
do Rio Grande do Sul, CEP 91501-970, Porto Alegre, RS, Brazil}

\begin{abstract}

We have analyzed the frequency and properties of the nuclear
activity in a sample of galaxies with circumnuclear rings and
spirals (CNRs). This sample was compared with a control sample of
galaxies with very similar global properties but without
circumnuclear rings. We discuss the relevance of the results in
regard to the AGN feeding processes and present the following
results: (i) bright companion galaxies seem not to be important
for the appearance of CNRs, which appear to be more related to
intrinsic properties of the host galaxies or to minor merger
processes; (ii) the proportion of strong bars in galaxies with an
AGN and a CNR is somewhat higher than the expected ratio of
strongly barred AGN galaxies from the results of Ho and coworkers;
(iii) the incidence of Seyfert activity coeval with CNRs is
clearly larger than the rate expected from the morphological
distribution of the host galaxies; (iv) the rate of Sy\,2 to Sy\,1
type galaxies with CNRs is about three times larger than the
expected ratio for galaxies without CNRs and is opposite to that
predicted by the geometric paradigm of the classical unified model
for AGNs, although it does support the hypothesis that Sy\,2
activity is linked to circumnuclear star formation.

\end{abstract}



\section{INTRODUCTION}

The unified standard model for active galactic nuclei stands on a
geometric paradigm, which implies that most of the observed
properties among the different kinds of objects arise from the
observer position and not from intrinsic properties of the host
environment or galaxy, in particular this statement is more clear
for the differences among Seyfert galaxies or among radio
galaxies.  This strong constraint makes important any
observational suggestion about systematic differences in the host
properties and has lead to a variety of statistical studies
without strong results.  The subject of AGN environment is still
open to discussion and clearly more studies and correlation
searches are needed.

In particular, the connection between nuclear activity, star
formation and infalling gas has received growing attention over
the past ten years, with nuclear bars and circumnuclear disk
instabilities being invoked as preferred mechanisms for removing
angular momentum from the gaseous fuel (e.g. Shlosman et al. 1989,
Heller \& Shlosman 1994, Maiolino et al. 2000).  The relationship
between the star-forming circumnuclear rings and ring-like
circumnuclear spirals (hereafter CNRs) and the resonances that
help to accumulate gas in these rings has been extensively
discussed from both the theoretical (e.g. Piner, Stone \& Teuben
1995, Wada \& Habe 1995) and the observational (e.g.
Storchi-Bergmann et al. 1996, D\'{\i}az et al. 1999) standpoints.
It has been claimed that bars and rings are more prevalent in
active and starburst galaxies than in otherwise normal objects
(Arsenault 1989).  Specifically, some studies have yielded
evidence that Seyferts have a preference for systems with global
inner or outer rings (Simkin et al. 1980, Arsenault 1989, Moles et
al. 1995, Hunt et al 1999). Notwithstanding, the subject of bars
is still controversial, bars seem to contribute significantly to
circumnuclear star formation, but without an evident relation with
nuclear ``non-stellar'' activity (Ho et al. 1996).

We have been using observational methods to study the mechanisms
related to nuclear activity and star formation in the central
regions of spiral galaxies.  In particular, our research program
has been focused in kinematical and dynamical detailed studies of
the central region of CNR galaxies (D\'{\i}az 2004). In the case
of NGC\,1241 (D\'{\i}az et al. 2003) we have shown the presence of
several perturbations in the hundred parsec scale linked to the
CNR where active star formation takes place with nuclear symmetry,
some of this perturbations being good candidates for gas angular
momentum removal and feeding of the Seyfert\,2 nucleus of
NGC\,1241. Our results encouraged us to perform a systematic
search for the presence of nuclear activity in galaxies with
circumnuclear rings or ring-like nuclear spirals (hereafter CNR
galaxies).

We have concentrated the present study in CNRs because they are
morphological structures which are radially well differentiated
from the active nucleus itself and other structures that can be
associated to the active nuclei, like outflows or extended
ionization regions. In disadvantage other circumnuclear star
formation features as, for example, nuclear bars and hot spots,
can be confused at low resolution, with AGN-related structures.
Besides, CNRs represent a defined stage in the secular evolution
of barred systems (e.g. Combes 2000) making more probable the
detection of any correlation with a defined AGN feeding stage. In
the present paper we describe how was statistically compared the
nuclear activity of galaxies of similar morphological type, with
and without CNRs.


\section{CATALOG AND ANALYSIS}

A list of 81 CNR bright galaxies (available on request) was
compiled by us, 64 of which were obtained from Buta \& Crocker
(1993) Catalog. Nine objects came from various authors: NGC\,1672,
Storchi-Bergmann et al. (1996); NGC\,1667 and NGC\,4151,
Kotilainen \& Ward (1997); NGC\,3032 and NGC\,7743, Erwin \&
Sparke (2002), NGC\,3516 and NGC\,3982, P\'erez-Ram\'{\i}rez et
al. (2000); NGC\,5327 and NGC\,5643, Laine et al. (1997). The
remaining eight objects came from our studies: NGC\,1241,
D\'{\i}az et al. (2003); NGC\,1566, Ag\"uero et al. (2004);
NGC\,300, Mrk 1066, NGC\,6221, NGC\,6300, NGC\,7479 and NGC\,7582,
D\'{\i}az (2004).  Most of the objects (64) come from
morphological studies of galaxies through CCD imaging surveys and
atlases searches, compiled or performed by Buta and Crocker
(1993). The ones added here were included if they have undoubtedly
defined rings from ground based observations and are nearby well
resolved features ($z<0.002$), without searching in specific AGN
studies. From the whole sample, we detected two ``interlopers´´
(NGC\,7469, Wilson et al. 1991; NGC\,1566, Ag\"uero et al. 2004)
originated in a specific paper on AGN study which reported a new
circumnuclear ring. We plotted the global properties of the sample
and they do not differ from that shown by Buta and Crocker (1993),
and there is no apparent difference between those corresponding to
circumnuclear rings and the comparison sample presented below, due
to the strong matching requirements.

It is intended to include most of the circumnuclear rings known
from global morphological searches (available in the literature
for $z<0.002$), and the number of objects is enough for some of
the trends found to be well over the N$^{1/2}$ threshold, which is
reported as uncertainty for each result.

The activity class of the 81 objects was obtained from the catalog
of V\'eron-Cetty \& V\'eron (2003) and the observations of nearby
galaxies by Ho et al. (1997).  The activity types compiled by
Veron-Cetty and Veron (1998) are almost coincident with those
independently reported in the deep spectral survey of Ho et al.
(1997), whenever the objects are common, which suggests that the
relative distribution of Seyfert types is not strongly dependent
on the two classification sources used here.

Of the 81 CNR galaxies which we studied, 60 are included in
Tully's (1988)``Catalog of Nearby Galaxies'', which was used as
source for the global properties. Hereafter we refer the complete
list of 81 objects as the full sample, and the selected 60, as the
partial sample. In order to asses any possible relationships
between nuclear and circumnuclear activity we made a list of
comparison galaxies, each one selected as the best match in
Tully's (1988) Catalog, based on the following criteria:

(i) the departure in B absolute magnitude from the CNR galaxy must
be $\Delta M_B < 0.3$;

(ii) the difference in corrected apparent sizes must be $\Delta
D_{25} < 0.2 \,D_{25}^{CNR}$;

(iii) the difference in projected real sizes must be $\Delta
R_{25} (kpc) < 0.4 \,R_{25}^{CNR} (kpc)$;

(iv) the departure in inclination from the CNR galaxy must be
$\Delta i < 9^o$;

(v) the departure in morphological type numerical code from the
CNR galaxy must be $\Delta T \leq 2$.

60 comparison galaxies were found without trace of CNR according
to the visual inspection in the 2MASS and the DSS2 archives (with
about 1\,arcsec resolution). Moreover, the selected matching
objects from Tully (1988) have the same distributions of
distances, inclinations and brightness, therein they have the same
detection probability that the hitherto known CNR galaxies. One
should not forget the possible presence of interlopers in the
comparison sample, but all of the comparison objects are bright
nearby galaxies and were not found reported as CNR galaxies.


\section{RESULTS}

\paragraph{Environment.}
The local density of bright galaxies around each object in the
partial and matched comparison sample (i.e. CNR or comparison
galaxy) from Tully's (1988) Catalog shows that (within the
uncertainty levels) there is no marked environmental effect
associated with the phenomenon of circumnuclear star formation in
disk galaxies (Figure\,1). The distributions of local densities
have peaks in $0.54\pm0.12$ and $0.62\pm0.14$ galaxies/Mpc$^{3}$
for the CNR and the comparison samples, respectively.

\paragraph{Strong Bars.}
We investigated the presence of optical bars in different subsets
of the partial sample and we found that $46\pm12$\% of 30 CNR+AGN
galaxies (Sy1, Sy2, LINER) and $42\pm8$\% of all (i.e. with and
without AGN) CNR galaxies were strongly barred, i.e. they are
classified as SB in the RC3 Catalog (de Vaucouleurs et al. 1991).
This proportion was higher (without overlapping of both N$^{1/2}$
uncertainties) than that in the sample of Ho et al. (1997), who
found that $20\pm4$\% of 129 AGN galaxies and $25\pm3$\% of 319
spiral galaxies with and without AGN were classified as SB in the
RC3 Catalog. Notwithstanding, in order to make a suitable
comparison, the morphological bias must be considered. Hence, we
rearranged the data with the same morphological type grouping used
by Ho and coworkers (who excluded the category S0/lenticular
objects), with the following results: $38\pm5$\% of 55 CNR
galaxies and $43\pm12$\% of 30 AGN+CNR ones, have strong bars,
compared with $24\pm6$\% and $20\pm8$\% respectively predicted
using Ho and co-workers results for each morphological type.
Consequently, we find some statistical excess (without overlapping
of both N$^{1/2}$ uncertainties) of strong bars in both CNR and
AGN+CNR galaxies, when compared with the spiral galaxies sample of
Ho and co-workers.

\paragraph{Incidence of AGNs.}
26 of the 81 objects ($32\pm6$\% of the full sample) were galaxies
Sy\,1 or 2, being this percentage unusually high, as shown in the
histogram of Figure\,2. For example, the value predicted
considering the analysis of Woltjer (1990) of the galaxies in the
Revised Shapley-Ames Catalog -by using the catalog of
V\'eron-Cetty \& V\'eron (1989) as AGN classification source-
would be $8\pm3$\% if we weight it for the distribution of
morphological types in our sample.  The excess is statistically
significative, without overlapping of almost three times both
N$^{1/2}$ uncertainties.

To check our results, we constructed a histogram (Figure 3) for
the partial sample and its comparison one. The predicted
percentage for the last one, weighted by morphological type was
$8\pm3$\%, which is close to the proportion (6/60) of Sy\,1 and
Sy\,2 galaxies found in this comparison sample and validates the
use of Woltjer (1990) data. For the objects with CNR, the value is
still clearly over the expected, with 24 Seyfert nuclei instead of
6.

\paragraph{Activity Classes.}
Following Maiolino \& Rieke (1995) we considered the class Sy1 as
the sum of classes Sy1+Sy1.2+Sy1.5. The ratio of Sy\,2 to Sy\,1
galaxies was 3:1 (Figures\,2 and 3) instead of the expected ratio
of about 1:1 for the distribution of morphological types in the
samples (considering Table\,5 in Woltjer 1990). This result being
statistical significant even for the 24 Seyfert galaxies in the
partial sample ($75\pm18$\% Sy\,2 against $25\pm10$\% Sy\,1
nuclei).

21 of the pairs CNR+Comparison galaxies in the partial sample are
included in the high quality spectral survey of Ho et al. (1997),
which is a complete survey and represents a more uniform AGN
classification source than V\'eron-Cetty \& V\'eron (2003). We
confirmed that all the mentioned trends in environment, bar
frequency, AGN incidence and Sy2:S1 ratio, are sustained, within
the uncertainties arisen in the lower number of objects of the
resulting sub-sample.

\section{DISCUSSION AND FINAL REMARKS}

\paragraph{Inclination.}
In order to assess the effect of inclination on the detection of
CNRs we constructed another comparison sample for the partial one,
but this time with free inclination value.  In general, CNRs are
assumed to be coplanar with the main disk and the result was as
expected, there being lack of galaxies with high inclinations in
the CNR partial sample due to a marked selection effect, because
of the fact that a highly inclined galactic disk precludes the
detection of CNRs.  This result also indicates that there is a
constraint to the fuelling mechanisms for central starbursts and
AGNs because the fact that the rings are coplanar with the global
disk implies that the angular momentum direction of the disk is
conserved by the infalling material down to the hundred parsecs
scale.  Any AGN feeding scenario for the central hundred parsecs
should take into account that between a few hundred parsec to a
few parsecs radii the material in the putative accretion disk and
the torus ceases to be influenced by the original direction of the
angular momentum of the global disk (see the ``Activity Classes''
Section, below).

\paragraph{Environment and Bars.}
The results shown in Figure\,1 suggest that CNRs are directly
associated with certain phenomena such as self instabilities,
minor mergers or the capture of giant HI clouds, unless they
belong to very late phase of interactions between galaxies, but
the work of Corbin (2000) suggests that this would be not the
case.  It should be noted that most of the galaxies in the full
sample had a global bar at optical or IR wavelengths (e.g. Buta \&
Crocker 1993) and about half of them were strongly barred, which
supports the widely accepted theory that bars are the main
mechanism which drives gas towards the central region of
non-interacting galaxies.

\paragraph{Incidence of AGNs.}
The observational bias must be assessed when collecting data
related to AGN galaxies.  For example, it could be argued that the
high rate of AGN reported here might be because the central region
of galaxies possessing Sy activity and the CNRs are more closely
studied than galaxies which do not have this features and there is
therefore a greater chance of detecting the remaining counterpart.
Not-withstanding, most of the CNR galaxies were discovered by
morphological searches (see Buta \& Crocker 1993 and Section 2).
After a careful search we have found 2 ``interlopers'' in the
sample which came from an AGN dedicated paper reporting the CNR
feature. This 2 AGN+CNRs do not alter at a level of significance
the results presented here, and even if a high interlopers number
of 10 CNRs discovered in AGN works was allowed in the sample, the
main trends reported on overabundance of AGNs still hold. In
particular, the ratio of Sy 2 to Sy 1 objects (3 times), is valid
even allowing for any bias toward AGN observations, unless one is
forced to assume that researchers would largely prefer to observe
Seyfert 2 galaxies and would consequently detect more frequently
their rings. Therein the correlation reported here must have some
statistical significance and presents an important restriction to
the models that describe the fuelling not only of AGN but also of
circumnuclear star formation.

\paragraph{Activity Classes.}
The high ratio of Sy\,2 to Sy\,1 galaxies seen by us is not in
accord with the classical unified model for AGNs, because
inclination of the host galaxy should have an important impact on
this model, unless one is forced to accept that the inclination of
the putative molecular torus has no relation, even in the
statistical sense, with the host global disk that eventually
provides the fuel with a specific angular momentum orientation. It
should be noted that there is an important bias towards face on
CNR hosts, so the observed incidence of Sy\,2 galaxies opposites
the expected incidence. In any case, the absence of correlation
between galaxy inclination and AGN orientation (e.g. Kinney et al.
2000) would impose important restrictions on the way the gas is
funnelled to the central engine in spiral galaxies.

Our results are in accordance with the observations of Malkan et
al. (1999), who determined that the morphology of the nuclear
region in the hundred parsec scale seems not to be an important
factor in the distribution of Sy classes, but that appearance of
circumnuclear filamentary dust is more related with Sy\,2
galaxies.   The intense star formation that occurs in most CNRs,
could provide the dust that, in some still undefined way, causes
the generally accepted obscuration of a Sy\,1 nucleus needed to be
observed as a Sy\,2. We expect soon to report the results of a
study on the correlation between the rate of star formation in
CNRs and the degree of AGN activity, such a study could help us
better understand the parallelisms between and co-evolution of
these interesting phenomena.

Hunt \& Malkan (1999) have found that outer rings but not bars are
abnormally frequent in Sy\,1 nuclei, while LINER galaxies appear
to have unusually high incidence of inner rings.  As these authors
suggested, a possible explanation might be to postulate that
LINERS and Sy nuclei are the same objects, but seen at different
evolutionary stages.  How does this scenario fit with our results?
Again the evolutionary solution must be recalled and becomes
reinforced:

i) LINER nuclei seem more coeval with inner rings (1.5 times the
ratio for normal spirals, Hunt \& Malkan 1999), which appear in
the first stage of bar evolution scenarios, between 2 and
5$\times10^8$\,yr to form (e.g. Combes \& Elmegreen 1993);

ii) Sy\,2 nuclei occur in large numbers in galaxies with
circumnuclear rings (4 times the ratio for normal galaxies, see
Figures 2 and 3 in this paper), which would require a large
central mass concentration fuelled by the bar (e.g. Combes 2000)
what in turn means that the bar is somewhat evolved and has had
enough time to sweep and remove the interstellar medium inside the
co-rotation radius, say more than 5$\times10^8$\,yr;

iii) Sy\,1 nuclei have 3-4 times more outer rings than normal
spirals and, as already pointed out by Hunt \& Malkan (1999), this
implies that Sy\,1 nuclei would be older enough, in an
evolutionary scheme, to coincide with the lifetime of outer rings
(3$\times10^9$\,yr), long after the bar has dissolved, due to the
expanding circularization of the orbits in the galaxy central
region (e.g. Combes 2000).

The trend pointed in paragraph (iii) would be equivalent to say
that Sy\,2 nuclei occur more at the stage of largest bar
dimensions (just before the dissolution), being this last idea
consistent with the results of Pogge (1989) and Maiolino et al.
(1997) which show that bar percentage in Sy\,2 galaxies appears
higher than in other AGN types.

The pointed correlation trends would be reinforced by the fact
that inner and circumnuclear rings can have lifetimes as short as
$10^8$\,yr, although these short time would preclude the detection
of larger fractions of the corresponding active counterparts.
Therefore, it seems plausible that the AGN evolutionary sequence
is from LINER to Sy\,2 and from this to Sy\,1, with the Sy\,2
class coeval with the ``dustier'' star formation era in the
bar-feeded evolution of the circumnuclear environment.

\section{Acknowledgements}

We are indebted to G. Carranza for his generous support. A
preliminary version of this work was presented to the Argentine
Astronomical Association in September of 2003. HD thanks the
brazilian institutions Conselho Nacional de Pesquisas (CNPq) and
Coordenadora para aperfei\c{c}oamento do Pessoal de Ensino
Superior (CAPES). This research is also partially supported by
brazilian grants MEGALIT/Millennium.






\clearpage

\begin{figure}
\includegraphics{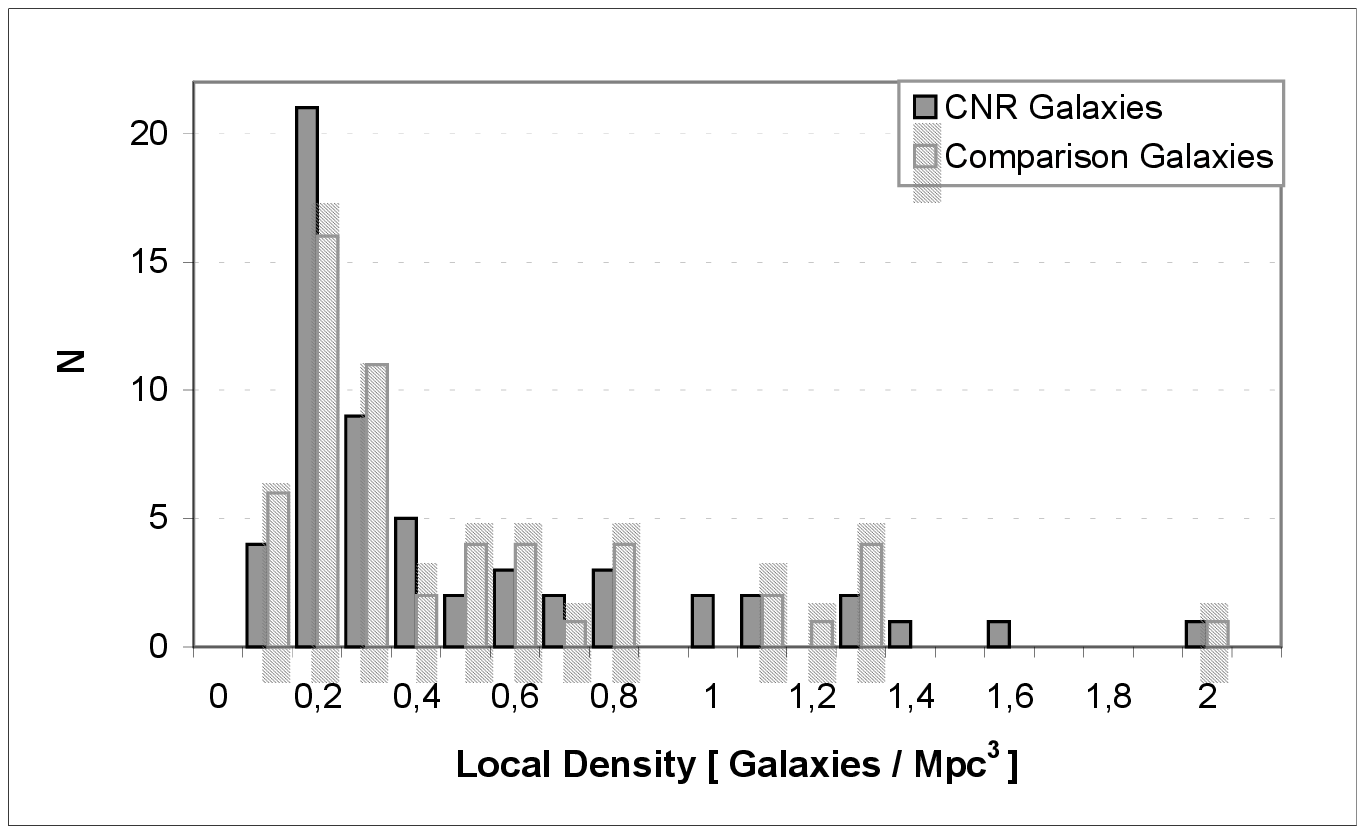} \vspace{15 cm} \caption{Local density distribution of
the CNR and comparison galaxies.} \label{figure1}
\end{figure}

\clearpage

\begin{figure}
\includegraphics{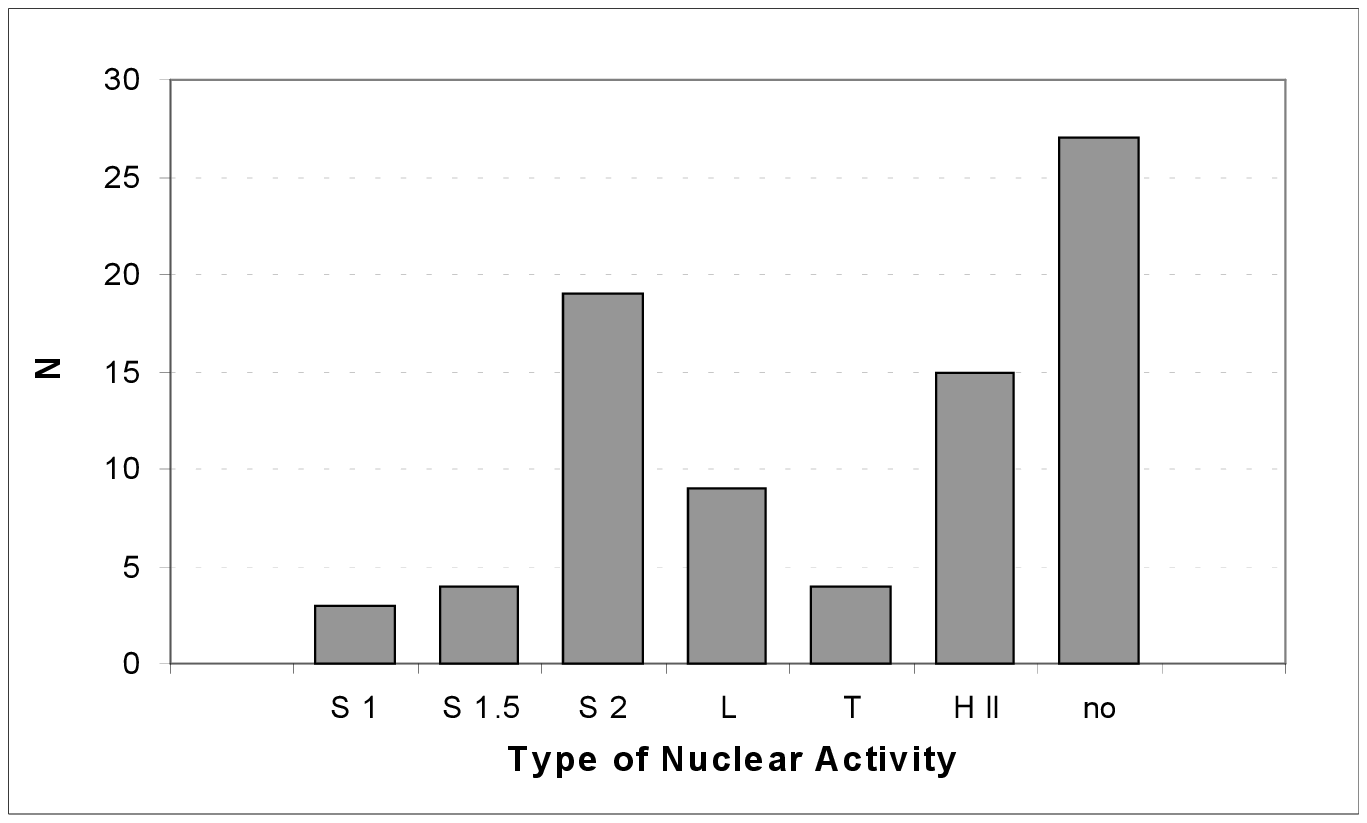} \vspace{15 cm} \caption{Level of nuclear activity of
the full sample of CNR galaxies. The sources of the activity type
are V\'eron-Cetty \& V\'eron (2003) and Ho et al. (1997). We have
separated the galaxies into categories as follows: S\,1-2
(Seyfert), L (LINER), T (transition object), HII (starburst or
normal HII region nucleus). Depending on each object, ``no'' means
that the object is not reported in these sources, or is reported
as not having line emission.} \label{figure2}
\end{figure}

\clearpage

\begin{figure}
\includegraphics{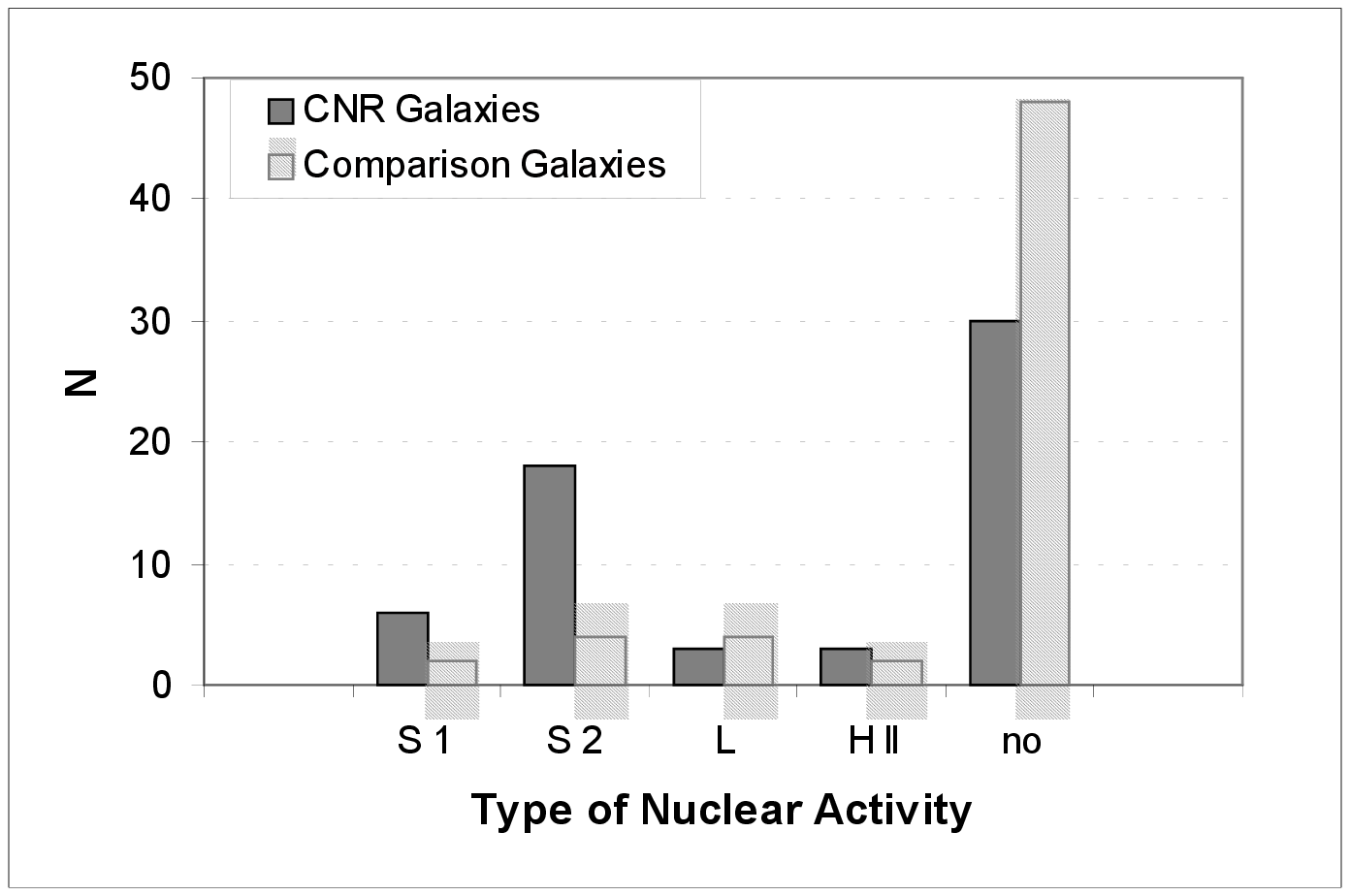} \vspace{15 cm} \caption{Level of nuclear activity in
the partial sample, both CNR and comparison galaxies. The source
of the activity type is V\'eron-Cetty \& V\'eron (2003). In this
Catalog, ``no'' means that the object is not reported in the
sources as AGN or strong HII nucleus.} \label{figure3}
\end{figure}

\clearpage

\end{document}